# QUAREP-LiMi: A community-driven initiative to establish guidelines for quality assessment and reproducibility for instruments and images in light microscopy

Glyn Nelson[1,x,*], Ulrike Boehm[2,x,*], Steve Bagley[3,x], Peter Bajcsy[4,x], Johanna Bischof[5,x], Claire M Brown[6,x], Aurélien Dauphin[7,x], Ian M Dobbie[8,x], John E Eriksson[9,x], Orestis Faklaris[10,x], Julia Fernandez-Rodriguez[11,x], Alexia Ferrand[12,x], Laurent Gelman[13,x], Ali Gheisari[14,x], Hella Hartmann[14,x], Christian Kukat[15,x], Alex Laude[1,x], Miso Mitkovski[16,x], Sebastian Munck[17,x], Alison J North[18,x], Tobias M Rasse[19,x], Ute Resch-Genger[20,x], Lucas C Schuetz[21,x], Arne Seitz[22,x], Caterina Strambio-De-Castillia[23,x], Jason R Swedlow[24,x], Ioannis Alexopoulos[25], Karin Aumayr[26], Sergiy Avilov[27], Gert-Jan Bakker[28], Rodrigo R Bammann[29], Andrea Bassi[30], Hannes Beckert[31], Sebastian Beer[32], Yury Belyaev[33], Jakob Bierwagen[34], Konstantin A Birngruber[35], Manel Bosch[36], Juergen Breitlow[37], Lisa A Cameron[38], Joe Chalfoun[4], James J Chambers[39], Chieh-Li Chen[40], Eduardo Conde-Sousa[41,42], Alexander D Corbett[43], Fabrice P Cordelieres[44], Elaine Del Nery[45], Ralf Dietzel[46], Frank Eismann[47], Elnaz Fazeli[48], Andreas Felscher[49], Hans Fried[50], Nathalie Gaudreault[51], Wah Ing Goh[52], Thomas Guilbert[53], Roland Hadleigh[29], Peter Hemmerich[54], Gerhard A Holst[55], Michelle S Itano[56], Claudia B Jaffe[57], Helena K Jambor[58], Stuart C Jarvis[59], Antje Keppler[60], David Kirchenbuechler[61], Marcel Kirchner[15], Norio Kobayashi[62], Gabriel Krens[63], Susanne Kunis[64], Judith Lacoste[65], Marco Marcello[66], Gabriel G Martins[67], Daniel J Metcalf[29], Claire A Mitchell[68], Joshua Moore[69], Tobias Mueller[70], Michael S Nelson[71], Stephen Ogg[72], Shuichi Onami[73], Alexandra L Palmer[74], Perrine Paul-Gilloteaux[75], Jaime A Pimentel[76], Laure Plantard[13], Santosh Podder[77], Elton Rexhepaj[78], Arnaud Royon[80], Markku A Saari[81], Damien Schapman[82], Vincent Schoonderwoert[83], Britta Schroth-Diez[84], Stanley Schwartz[85], Michael Shaw[86], Martin Spitaler[87], Martin T Stoeckl[88], Damir Sudar[89], Jeremie Teillon[90], Stefan Terjung[21], Roland Thuenauer[91], Christian D Wilms[29], Graham D Wright[52], and Roland Nitschke[92,x,+]

[1]Bioimaging Unit, Newcastle University, Newcastle upon Tyne, NE4 5PL, UK
[2]Janelia Research Campus, Howard Hughes Medical Institute, Ashburn, VA 20147, USA
[3]Visualisation, Irradiation & Analysis, Cancer Research UK Manchester Institute, Alderley Park, Macclesfield, UK
[4]National Institute of Standards and Technology, Gaithersburg, MD 20899, USA
[5]Euro-BioImaging, Heidelberg, 69117, Germany
[6]Advanced BioImaging Facility (ABIF), McGill University, Montreal, Quebec, H3G 0B1, Canada
[7]Unité Génétique et Biologie du Développement U934, PICT-IBiSA, Institut Curie/Inserm/CNRS/PSL Research University, Paris, 75005, France
[8]Department of Biochemistry, University of Oxford, Oxford, Oxon, OX1 3QU, UK
[9]Turku Bioscience Centre, Euro-Bioimaging ERIC, Turku, 20520, Finland
[10]Biocampus, CNRS UAR 3426, Montpellier 34293, France
[11]Centre for Cellular Imaging, Sahlgrenska Academy, University of Gothenburg, Gothenburg, 40530, Sweden
[12]Imaging Core Facility, Biozentrum, University of Basel, Basel, 4056, Switzerland
[13]Friedrich Miescher Institute for Biomedical Research, Basel, 4058, Switzerland
[14]Light Microscopy Facility, CMCB Technology Platform, TU Dresden, Dresden, 01307, Germany

[15]FACS & Imaging Core Facility, Max Planck Institute for Biology of Ageing, Cologne, 50931, Germany
[16]Light Microscopy Facility, Max Planck Institute of Experimental Medicine, Goettingen, 37075, Germany
[17]VIB BioImaging Core & VIB-KU Leuven Center for Brain and Disease Research & KU Leuven Department for Neuroscience, Leuven, Flanders, 3000, Belgium
[18]The Rockefeller University, New York, NY 10065, USA
[19]Scientific Service Group Microscopy, Max Planck Institute for Heart and Lung Research, Bad Nauheim, 61231, Germany
[20]Division Biophotonics, Federal Institute for Materials Research and Testing, Berlin, 12489, Germany
[21]European Molecular Biology Laboratory, Advanced Light Microscopy Facility, Heidelberg, 69117, Germany
[22]Faculty of Life Sciences, Ecole Polytechnique Fédérale de Lausanne, Lausanne, Vaud, 1015, Switzerland
[23]Program in Molecular Medicine, University of Massachusetts Medical School, Worcester, MA 01605, USA
[24]Divisions of Computational Biology and Gene Regulation and Expression, School of Life Sciences, University of Dundee, Dundee, UK
[25]General Instrumentation - Light Microscopy Facility, Radboud University, Faculty of Science, Nijmegen, 6525, Netherlands
[26]BioOptics Facility, IMP - Research Institute of Molecular Pathology, Vienna, 1030, Austria
[27]Max Planck Institute of Immunobiology and Epigenetics, Freiburg, 79108, Germany
[28]Department of Cell Biology (route 283), Radboud Institute for Molecular Life Sciences, Nijmegen, 6525GA, Netherlands
[29]Scientifica Ltd, Uckfield, East Sussex, TN22 1QQ, UK
[30]Dipartimento di Fisica, Politecnico di Milano, Milan, 20133, Italy
[31]Microscopy Core Facility, Universität Bonn, Medizinische Fakultät, Bonn, 53127, Germany
[32]Hamamatsu Photonics GmbH, Herrsching, 82211, Germany
[33]Microscopy Imaging Center, University of Bern, Bern, 3012, Switzerland
[34]AHF analysentechnik AG, Tuebingen, 72074, Germany
[35]TOPTICA Photonics AG, Graefelfing, 82166, Germany
[36]Fac. Biology, Prevosti's building, Universitat de Barcelona, Barcelona, Spain
[37]PicoQuant, Berlin, 12489, Germany
[38]Light Microscopy Core Facility; Biology, Duke Univeristy, Durham, NC 27708, USA
[39]Institute for Applied Life Sciences, University of Massachusetts, Amherst, MA 01003, USA
[40]Pathware, Seattle, WA 98121, USA
[41]i3S - Instituto de Investigação e Inovação em Saúde, Universidade do Porto, 4169-007 Porto, Portugal
[42]INEB - Instituto de Engenharia Biomédica, Universidade do Porto, 4169-007 Porto, Portugal
[43]Department of Physics and Astronomy, University of Exeter, Exeter, EX4 4QL, UK
[44]Bordeaux Imaging Center, Bordeaux, Nouvelle Aquitaine, 33077, France
[45]BioPhenics High-Content Screening Laboratory (PICT-IBiSA), Translational Research Department, Institut Curie - PSL Research University, Paris, 75248, France
[46]Omicron-Laserage Laserprodukte GmbH, Rodgau, 63110, Germany
[47]Carl Zeiss Microscopy GmbH, Jena, 07745, Germany
[48]University of Turku, Turku, 20520, Finland
[49]Coherent Lasersysems GmbH & Co.KG, Luebeck, 23569, Germany
[50]Light Microscope Facility, German Center for Neurodegenerative Diseases (DZNE), Bonn, 53127, Germany
[51]Allen Institute for Cell Science, Seattle, WA 98109, USA
[52]A*STAR Microscopy Platform, Research Support Centre, Agency for Science, Technology and Research, Singapore, 138648, Singapore
[53]Institut Cochin, INSERM (U1016), CNRS (UMR 8104), Université de Paris (UMR-S1016), Paris, 75014, France
[54]Core Facility Imaging, Leibniz Institute on Aging, Jena, 07745, Germany
[55]Research & Science, PCO AG, Kelheim, 93309, Germany
[56]Neuroscience Microscopy Core, University of North Carolina, Chapel Hill, NC 27599-7250, USA
[57]Lumencor, Inc., Beaverton, OR 97006, USA
[58]Mildred-Scheel Nachwuchszentrum, Universitätsklinikum Carl Gustav Carus, TU Dresden, Dresden, 01307, Germany
[59]Prior Scientific Instruments Limited, Cambridge, Cambridgeshire, CB21 5ET, UK
[60]EMBL Heidelberg, Global BioImaging, Heidelberg, 69117, Germany
[61]Northwestern, Chicago, IL 60611, USA

[62]RIKEN, Wako, Saitama, 351-0198, Japan
[63]Bioimaging Facility, Institute of Science and Technology Austria, Klosterneuburg, 3400, Austria
[64]Biology/Chemistry, University Osnabrueck, Osnabrueck, 49080, Germany
[65]MIA Cellavie Inc., Montreal, Quebec, H1K 4G66, Canada
[66]Inst. of Systems, Molecular & Integrative Biology, University of Liverpool, Liverpool, Merseyside, L697ZB, UK
[67]Instituto Gulbenkian de Ciencia & Faculdade de Ciencias, Univ. Lisboa., Oeiras, 2780-156, Portugal
[68]Warwick Medical School, University of Warwick, Coventry, West Midlands, CV4 7AL, UK
[69]School of Life Science, University of Dundee, Walluf, 65396, Germany
[70]Gregor Mendel Institute of Molecular Plant Biology (GMI), Vienna, 1030, Austria
[71]City of Hope, Duarte, CA 91010, USA
[72]Medical Microbiology & Immunology, University of Alberta, Edmonton, Alberta, T6G 2E1, Canada
[73]RIKEN Center for Biosystems Dynamics Research, Kobe, Hyogo, 650-0047, Japan
[74]Advanced Light Microscopy, The Francis Crick Institute, London, NW1 1AT, UK
[75]Université de Nantes, CHU Nantes, Inserm, CNRS, SFR Santé, Inserm UMS 016, CNRS UMS 3556, F-44000 Nantes, France
[76]Instituto de Biotecnología, Universidad Nacional Autónoma de México, Cuernavaca, Morelos, 62210, Mexico
[77]Microscopy Facility, Biology, Indian Institute of Science Education and Research Pune, Pune, 411008, India
[78]Sanofi Aventis, Chilly-Mazarin, Essone, 91380, France
[80]Argolight, Pessac, 33600, France
[81]Turku Bioscience Centre, University of Turku and ˚ Abo Akademi University, Turku, 21520, Finland
[82]Normandie univ, UNIROUEN, INSERM, PRIMACEN, 76000 Rouen, France
[83]Scientific Volume Imaging bv, Hilversum, Noord-Holland, 1213VB, Netherlands
[84]Light Microscopy Facility, Max Planck Institute of Molecular Cell Biology and Genetics, Dresden, 01307, Germany
[85]Nikon Instruments Inc. ISO Consultant, Melville, NY 11747, USA
[86]National Physical Laboratory, Teddington, Middlesex, TW11 0LW, UK
[87]Imaging Facility, Max Planck Institute of Biochemistry, Martinsried, Munich, 82152, Germany
[88]Bioimaging Center, University of Konstanz, Konstanz, 78464, Germany
[89]Quantitative Imaging Systems, Portland, OR 97209, USA
[90]Bordeaux Imaging Center, Université de Bordeaux, Bordeaux, Gironde, 33076, France
[91]Technology Platform Microscopy and Image Analysis, Heinrich Pette Institute, Leibniz Institute for Experimental Virology, Hamburg, 20251, Germany
[92]Life Imaging Center and BIOSS Centre for Biological Signaling Studies, Albert-Ludwigs-University Freiburg, Freiburg, 79104, Germany
[*]these authors contributed equally to this work
[x]these authors are active members of QUAREP-LiMi's working group 8 and contributed significantly to the realization of this paper
[+]Corresponding author. Email: Roland.Nitschke@biologie.uni-freiburg.de

# ABSTRACT

A modern day light microscope has evolved from a tool devoted to making primarily empirical observations to what is now a sophisticated, quantitative device that is an integral part of both physical and life science research. Nowadays, microscopes are found in nearly every experimental laboratory. However, despite their prevalent use in capturing and quantifying scientific phenomena, neither a thorough understanding of the principles underlying quantitative imaging techniques nor appropriate knowledge of how to calibrate, operate and maintain microscopes can be taken for granted. This is clearly demonstrated by the well-documented and widespread difficulties that are routinely encountered in evaluating acquired data and reproducing scientific experiments. Indeed, studies have shown that more than 70% of researchers have tried and failed to repeat another scientist's experiments, while more than half have even failed to reproduce their own experiments[1]. One factor behind the reproducibility crisis of experiments published in scientific journals is the frequent underreporting of imaging methods caused by a lack of awareness and/or a lack of knowledge of the applied technique[2]. Whereas quality control procedures for some methods used in biomedical research, such as genomics (e.g., DNA sequencing, RNA-seq) or cytometry, have been introduced (e.g. ENCODE[3]), this issue has not been tackled for optical microscopy instrumentation and images. Although many calibration standards and protocols have been published, there is a lack of awareness and agreement on common standards and guidelines for quality assessment and reproducibility[4].

In April 2020, the QUality Assessment and REProducibility for Instruments and Images in Light Microscopy (QUAREP-LiMi) initiative[5] was formed. This initiative comprises imaging scientists from academia and industry who share a common interest in achieving a better understanding of the performance and limitations of microscopes and improved quality control (QC) in light microscopy. The ultimate goal of the QUAREP-LiMi initiative is to establish a set of common QC standards, guidelines, metadata models, and tools, including detailed protocols, with the ultimate aim of improving reproducible advances in scientific research.

This White Paper 1) summarizes the major obstacles identified in the field that motivated the launch of the QUAREP-LiMi initiative; 2) identifies the urgent need to address these obstacles in a grassroots manner, through a community of stakeholders including, researchers, imaging scientists[6], bioimage analysts, bioimage informatics developers, corporate partners, funding agencies, standards organizations, scientific publishers, and observers of such; 3) outlines the current actions of the QUAREP-LiMi initiative, and 4) proposes future steps that can be taken to improve the dissemination and acceptance of the proposed guidelines to manage QC.

To summarize, the principal goal of the QUAREP-LiMi initiative is to improve the overall quality and reproducibility of light microscope image data by introducing broadly accepted standard practices and accurately captured image data metrics.

## Preface

The QUality Assessment and REProducibility for Instruments and Images in Light Microscopy (QUAREP-LiMi) initiative[a] aims at convening the light microscopy community with the explicit purpose of reaching a broad consensus concerning Quality Control and Quality Assessment guidelines for optical microscopy to be adopted worldwide. For the purposes of this discussion, by "light microscopy community," we refer to everyone working directly or indirectly with light microscopes and image data, independent of the specific microscope design or configuration. Although we aim to satisfy the entire community's requirements and views, we cannot claim sufficient diversity or coverage of the community for complete representation. Rather, this White Paper is the first of a series that will report our ongoing progress towards achieving the stated goals of QUAREP-LiMi. While the work of QUAREP-LiMi aims at developing recommendations and guidelines that can be easily extended across disciplines (both physical and life sciences), for the sake of simplicity, the discussion is currently restricted to applications and examples drawn mainly from biology. Although our current efforts focus on establishing guidelines for widefield and confocal optical microscopes, we are keen to extend the breadth of our work subsequently to cover other light-microscopy-based imaging modalities.

## Background

### Current Situation

Since their introduction in the early 17th Century, microscopes have transitioned from basic, qualitative image-collecting tools to sophisticated instruments capable of automatically acquiring information-rich images that are further processed via advanced image processing and analysis steps to extract quantitative information about the underlying science. The robustness of the conclusions that we make from these observations will depend upon the reproducibility of the samples and the microscope system used to image them. Importantly, each instrument's technical characteristics need to be fully understood and documented to permit valid interpretation of imaging data. To enable the reliable and reproducible extraction of quantitative information, microscopes, including advanced widefield and confocal instruments, must therefore be well described, maintained, calibrated, and in essence 'quality controlled' (Figure 1).

[a] https://quarep.org/

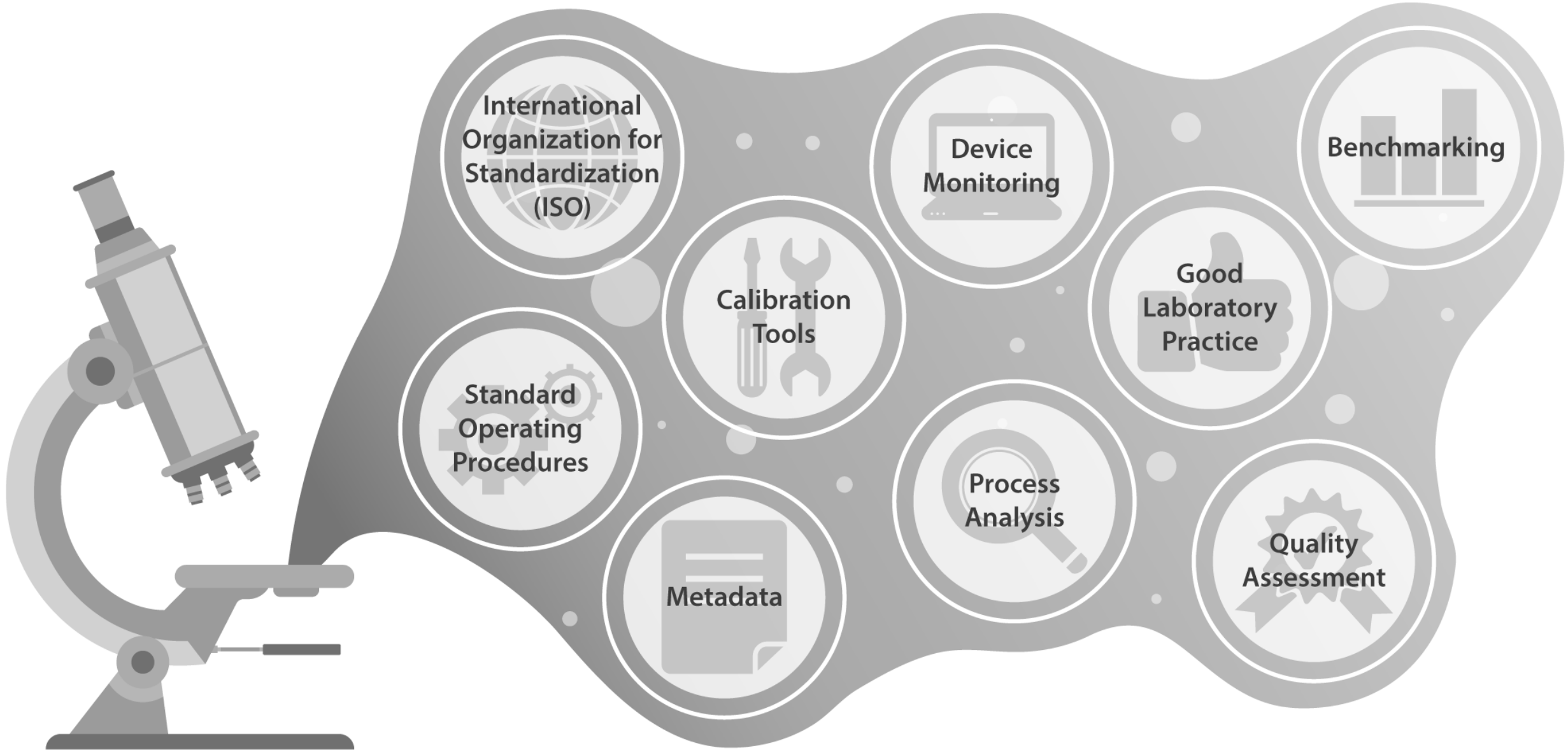

**Figure 1. Acquiring imaging data that is both quantifiable and reproducible involves a myriad of factors, few of which are acknowledged or accurately recorded.** Intimate knowledge of the composition and performance of a system is essential for reproducibility. However, performance measurements may be tricky, and require specific protocols, tools, samples, training, and data analysis methods. In order to help microscope users to assess and judge the performance of their systems properly, the community must agree on and publish guidelines and benchmarks.

Unfortunately, the requirement for robust microscope and image quality assessment is not commonly recognized amongst the scientific community, leading to the infrequent application of appropriate QC procedures. This is due to several barriers:

1. The lack of widely adopted community-wide guidelines and standards for light microscopy documentation and QC;
2. The limited willingness of the community's stakeholders (researchers, funders, and scientific publishers)[7] to enforce existing guidelines and standards;
3. Insufficient training of microscope users on the complexity of performing quantitative imaging and on guidelines and standards for quality assessment and reproducibility.

As a result, rigor and reproducibility are limited, the reliability of quantitative analysis is severely impacted, and the confidence in published data becomes eroded.

Encouragingly, the awareness of the importance of QC and reproducibility in light microscopy has gained traction in recent years, both within the scientific community and amongst funders (e.g., National Institutes of Health (NIH), European Research Council (ERC)) as well as within newly launched bioimaging networks (e.g., Max Planck BioImaging Core Unit Network[b]). More active steps have been taken within several microscopy initiatives such as Global Bioimaging (GBI[c]), Euro-BioImaging ERIC (European Research Infrastructure Consortium)[d], the Royal Microscopical Society (RMS[e]), German BioImaging-Gesellschaft für Mikroskopie und Bildanalyse (GerBI-GMB[f]), BioImaging North America (BINA[g]) and the RT-MFM technological network (Microscopie photonique de Fluorescence Multidimensionnelle[h]). GBI has published an overview of the current landscape for quality assurance and data management in imaging facilities[8,9], including recommendations for QC. This document also highlights multiple aspects concerning image data standardization, management, and publication, such as the definition of image and microscopy metadata guidelines and data models[10–13] and the need to provide open access to all raw data for accepted manuscripts (e.g., Image Data Resource and BioImage Archive)[14–17], which are being addressed both within QUAREP-LiMi (see Working Group 7 – Metadata) and by others[8–11] within the imaging community. The Euro-Bioimaging ERIC includes the independent assessment of QC measures and implementation in the ongoing evaluation of existing Nodes and during the application process for new imaging Node candidates. Finally, BINA, the RMS, GerBI-GMB, and the RT-MFM are all actively engaged in tackling QC and reproducibility issues via dedicated working groups ("QC and Data Management (QC-DM[i]); "QC Focussed Interest Group"[j]; "Quality Assessment for Instruments & Facilities"[k]; and "Metrological Measurements on Microscopy," respectively).

[b] https://www.bioimagingnet.mpg.de/aims [c] https://www.globalbioimaging.org/ [d] https://www.eurobioimaging.eu/ [e] https://www.rms.org.uk/
[f] https://www.gerbi-gmb.de/ [g] https://www.bioimagingna.org/ [h] http://rtmfm.cnrs.fr/

## Current Approaches

Despite the importance of individual local efforts, they prove insufficient to overcome the global challenges associated with QC in light microscopy. In the following section, we highlight a few of these approaches and discuss why they are unable to completely overcome existing hindrances individually.

*Quality Control procedures adopted by individual core facilities and laboratories*

To tackle common QC issues, many core facilities and laboratories regularly perform maintenance and various QC tests of their instruments. However, the nature and frequency of the performed tests vary greatly, depending on the priorities set by researchers, imaging facility staff, and their institution. A survey initiated by the European Light Microscopy Initiative (ELMI) in 2019[l] highlighted that numerous core facilities and labs already perform QC, but a considerable percentage does not at all (Figure 2a). Likewise, there was wide variation in the respondents' choice of tools, making any comparison and reproducibility of QC results between equipment difficult (Figure 2b).

*Guidelines by the International Organization for Standardization (ISO)*

The International Organisation for Standardization (ISO[m]) has created standards for brightfield microscopy[18,19] and, more recently, for confocal microscopy[20]. These ISO standards provide researchers with directions as to what should be measured and tested. Nevertheless, there is little information describing *how* key measurements should be made within these documents, using *which* samples and tools, and with *what frequency*.

*Tools and protocols by the community for the community*

Several individual groups have published methods and software tools to streamline and automate microscope QC procedures (e.g.[21–26] and recently reviewed in[27,28]). In addition, several open-source software tools that provide different degrees of automation for different microscopy calibration tasks have been developed and made available both as ImageJ-based macros and plugins (e.g., NoiSee[29], MetroloJ[30], ConfocalCheck[31], AutoQC[32], PSFj[33] and MIPs for PSFs[34], SIMcheck[35]) and standalone web applications (e.g., PyCalibrate[n]). Finally, international endeavors involving the global community were carried out and published by the Association of Biomolecular Resource Facilities (ABRF)[36,37]. These efforts provide both valuable results and metrics that can be saved locally and archived individually. However, these undertakings were affected by significant variations between individual groups. Moreover, they are neither comprehensive nor address standardization of metadata capture, and they are not fully aligned with the recommendations put forth by standards organizations such as the ISO.

*Regulations and guidelines imposed by third parties*

Besides the ISO, funding agencies, scientific publishers, and community organizations (e.g., GerBI-GMB, RMS, BINA, RT-MFM) often furnish QC guidelines for light microscopy. However, these guidelines are not exhaustive, are often issued in isolation, and are not accepted by the principal constituents of the imaging community (including commercial microscope manufacturers and the broader scientific community).

All of these approaches share a similar set of limitations: 1) they are currently adopted voluntarily and are therefore unenforceable; 2) they are often targeted at highly trained imaging facility staff and are often not accessible to less expert, non-facility microscope users, and custodians in individual laboratories; 3) they are limited in scope and therefore do not guarantee proper QC and reproducibility, and 4) they are not standardized and therefore show significant variability. The reasons for this are several-fold. Firstly, there is a lack of agreement regarding the recommended standard samples, tools, protocols, and metrics to be measured. Hence, there is a pressing need for the community to agree on what should be measured for each hardware component (e.g., laser, camera, or objective lens) and calibration procedure (i.e., optical, intensity, and mechanical calibration), which tools and samples should be used, and the frequency of QC measurements for different metrics. Secondly, a commonly cited reason for minimizing or avoiding microscope QC is the lack of appropriate time and resources (i.e., personnel, machine time, and required hardware) afforded to microscope custodians to perform the appropriate tests, the downstream analysis, and the compilation of the results across time accurately and systematically. Finally, most protocols/methods currently being performed are marred by high variability of the measured values, almost entirely due to the lack of automation.

[i] https://www.bioimagingna.org/qc-dm-wg [j] https://www.rms.org.uk/network-collaborate/focussed-interest-groups/quality-control.html
[k] https://www.gerbi-gmb.de/WG1 [l] https://lic-machform.vm.uni-freiburg.de/view.php?id=59721 [m] https://www.iso.org/ [n] https://www.psfcheck.com/

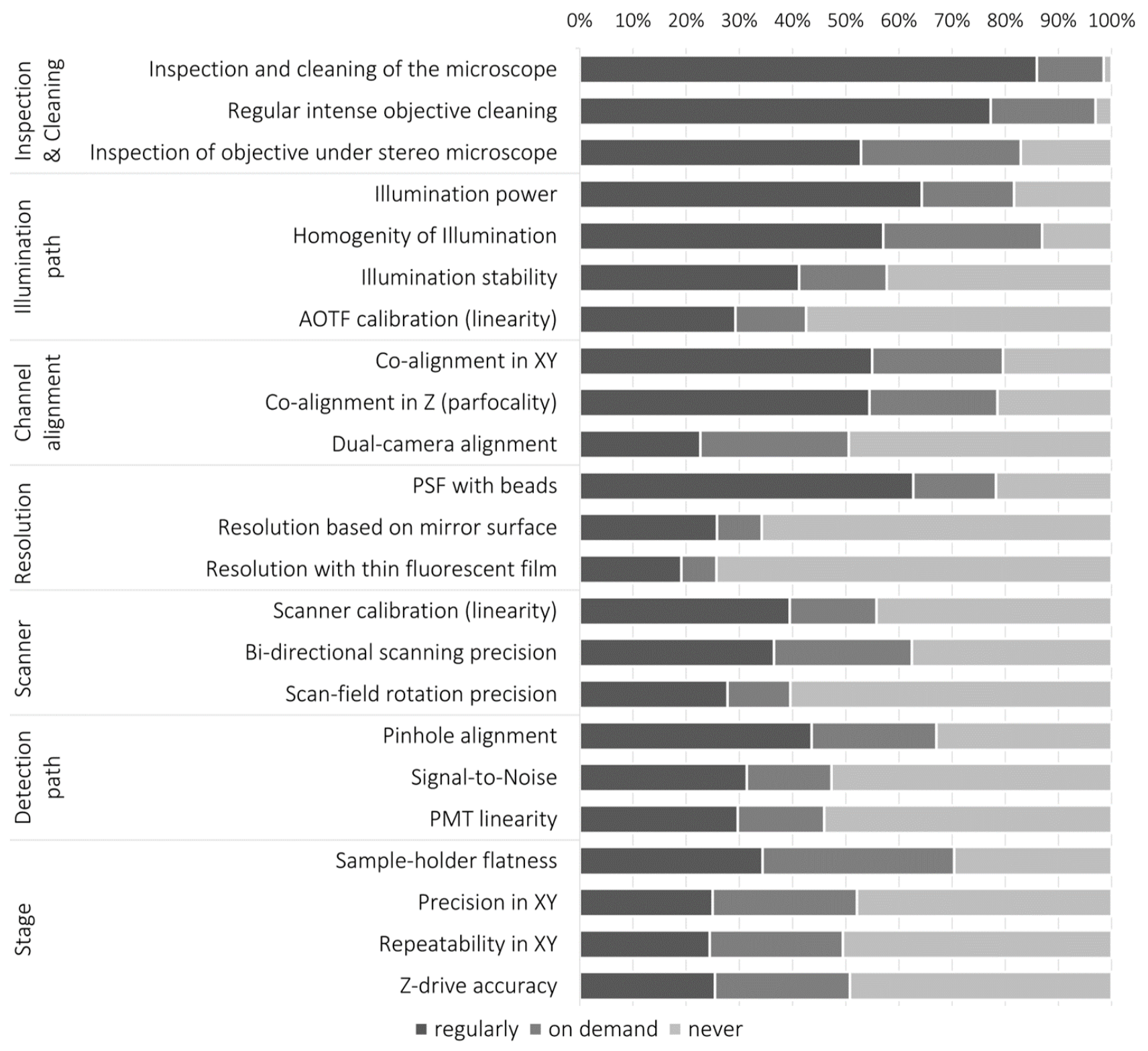

**(a)** Frequency of Quality Checks at Microscopy Facilities.

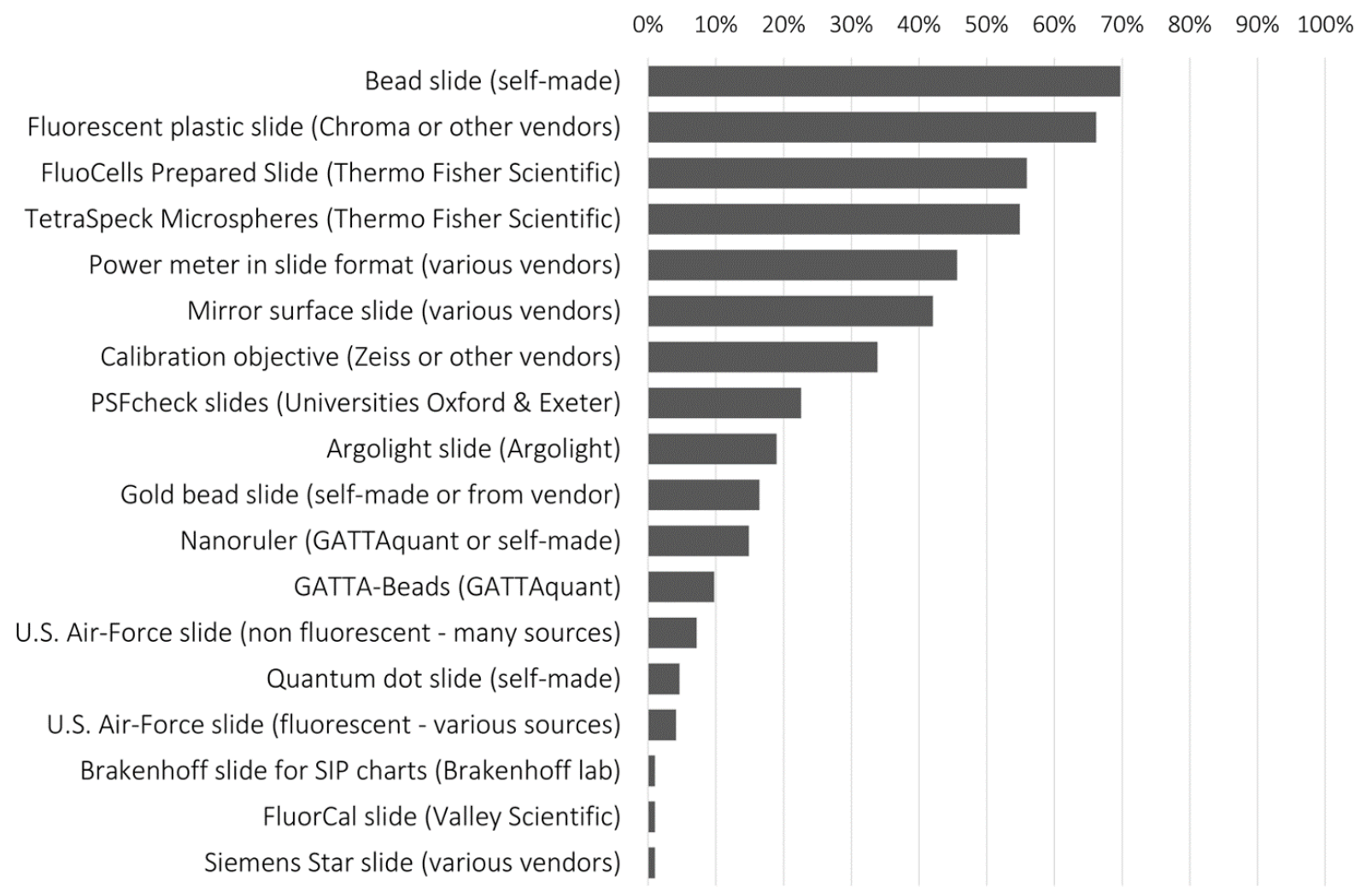

**(b)** Tools used for Performance Evaluation in Light Microscopy (ordered according to popularity).

**Figure 2. Frequency and type of tests performed in core facilities.** Prior to a session on Quality Control organized during the Microscopy Facility Day at the 2019 ELMI meeting, light microscopy core facility representatives around the globe were asked to complete a survey about the type and the frequency of tests performed in their facility. The link to the survey was opened in June 2019, two weeks before the meeting, and sent to all registered participants. It was also advertized multiple times on various international microscopy forums. Reminders were also sent to participants after the meeting and responses were subsequently collected until February 2020. The histograms in panels (a) and (b) summarize the responses from almost 200 facilities in a simplified manner. Panel (a) displays how often different quality checks are performed; the x axis represents in percentage the respective frequency categories, namely regularly, on demand, and never. Panel (b) highlights which tools are used for performance evaluation of light microscopes; the x axis represents the percentage of respondents using the indicated QC tool.

## Proposed Community-Driven Approach

Following a discussion at the 2019 conference of the European Light Microscopy Initiative (ELMI 2019), members of the GerBI-GMB and RT-MFM networks launched a shared strategy to build a community consensus on QC measurements. This initial initiative rapidly integrated with similar efforts being conducted by the BINA QC-DM working group[o] and the RMS QC focussed interest group[p]. Shortly after, the publication of the Confocal ISO 21073[20] provided the scientific community with an agreed minimal set of tests that should be performed to assess the performance of confocal microscopes. Based upon this publication, a methodology manual was drafted to describe how the ISO-recommended QC metrics could be obtained in practice[38] and presented to participants representing academia, industry, and governmental standardization bodies at a meeting held on April 28th, 2020. This led to the formal establishment of QUAREP-LiMi[q] (coordinated by R. Nitschke, Gerbi-GMB). QUAREP-LiMi is a grass-root global community that is open to individuals from academia, industry, government, funding agencies, and scientific journals from around the world with interest in improving QC in light microscopy. As a testament to the timeliness of this strategy, QUAREP-LiMi quickly grew from 49 initial participants to 184 (updated 05/01/2021) individuals (at the time of writing) from 20 countries (Figure 3). Compared to earlier approaches (see section "Current Approaches"), the QUAREP-LiMi initiative is specifically designed to work in a completely transparent and open manner to foster ground-up participation from around the globe and ownership by all members of the community. By taking into account all existing approaches and recommendations, the specific goal is to produce a consensus around shared, binding QA and QC guidelines and specifications for the scientific community and corporate partners. Furthermore, QUAREP-LiMi will work with both journals and funders to encourage stakeholders to adopt and enforce these standards.

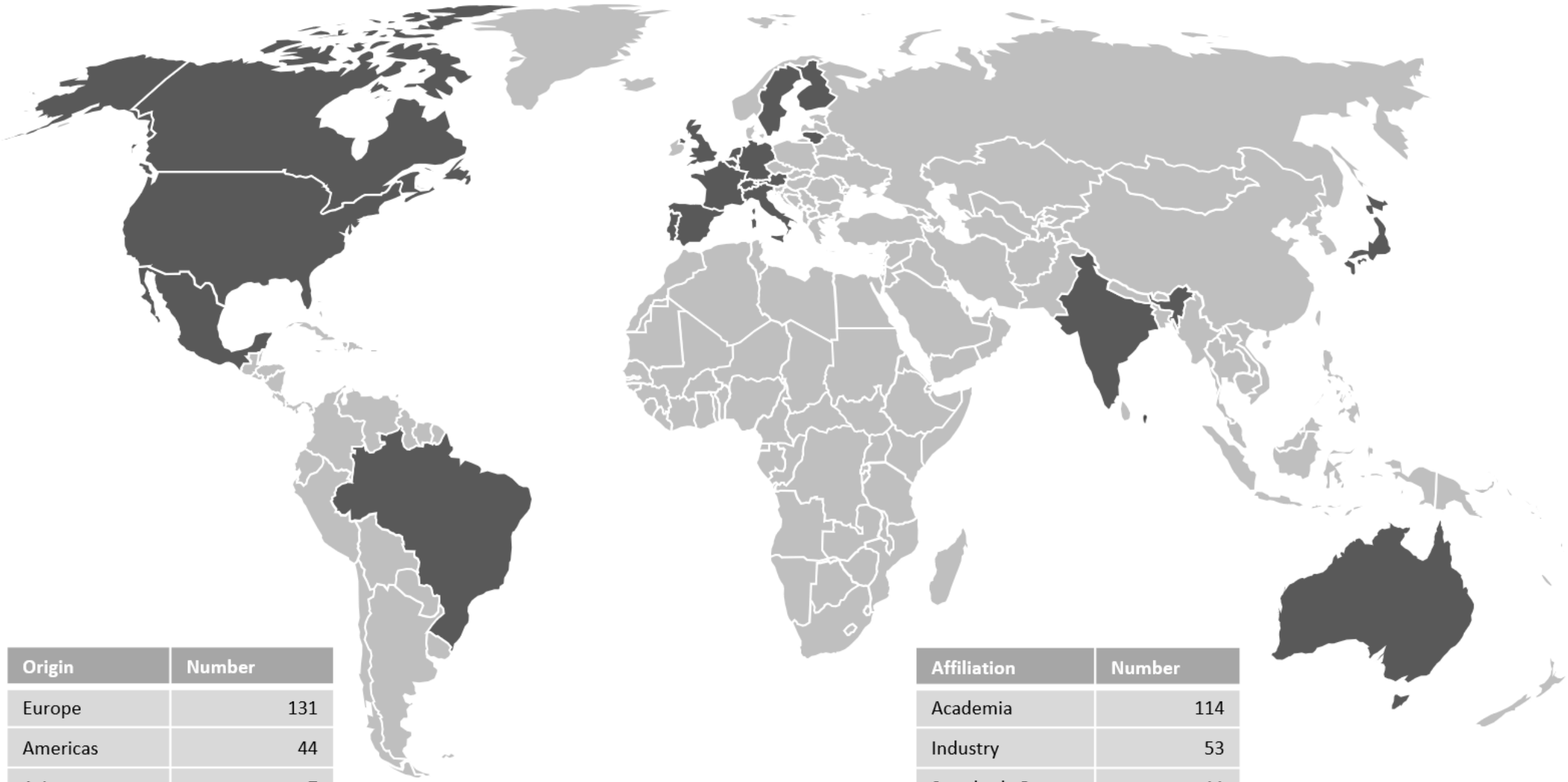

| Origin | Number |
|---|---|
| Europe | 131 |
| Americas | 44 |
| Asia | 7 |
| Australia | 2 |
| Total | 184 |

| Affiliation | Number |
|---|---|
| Academia | 114 |
| Industry | 53 |
| Standards Org. | 11 |
| Others | 6 |
| Total | 184 |

**Figure 3. Summary of current QUAREP-LIMi participants according to their origin and affiliation** *(updated: January 5, 2021).*

[o] https://www.bioimagingna.org/qc-dm-wg
[p] https://www.rms.org.uk/network-collaborate/focussed-interest-groups/quality-control.html
[q] https://quarep.org/

## Key Beneficiaries

The deliverables from QUAREP-LiMi will benefit several groups related to light microscopy, from image data acquisition all the way to image data processing, presentation, sharing, and reuse.

1. *Research scientists and imaging scientists* will take advantage of the harmonization and simplification of the QC procedures, facilitation of QC capture and storage, and clear interpretation of QC results to better understand how the performance of their microscope impacts the interpretation of scientific results.

2. *Scientific publishers and the general public* will profit from an overall enhanced trust in the value and reproducibility of scientific publications, resulting from the publication of full descriptions (i.e., Material and Methods sections and microscopy metadata) of the technical make-up of microscopes and of performance metrics to accompany raw image data.

3. *Funding bodies* will benefit from the planned improvement of QC practices that will undoubtedly improve image data reliability, reproducibility, and openness, increasing overall the value and quality of scientific output and opening the way to truly FAIR data[39]. Besides, the improved likelihood of data reuse towards novel discoveries will significantly impact taxpayer funds' efficient use.

4. *Core imaging facilities and their users, as well as non-facility microscope custodians and users,* will profit from heightened confidence in the accuracy of their image data and the suitability of performed QC measurements to answer their scientific questions. Improving, prioritizing, and streamlining the various required QC procedures will ease the time burden, facilitate the comparison of experimental data[40] and improve the communication between imaging facility staff and commercial manufacturers based on common vocabulary, tools, and protocols. As an added advantage, community standard procedures will allow users to submit standardized performance metrics with their published imaging data, which will greatly facilitate the interpretation, reproducibility, and re-use of the results.

5. *Commercial microscope and system component manufacturers* will be able to take advantage of the availability of time-stamped and standardized microscope metrics to identify common microscope performance and quality issues, such as identifying faulty parts. They will also be able to utilize the metrics for future developments and the continued improvement of hard- and software products.

## Initiated Steps and Dissemination

Although initiated around the Confocal Microscope ISO 21073 standard[19], the scope of QUAREP-LiMi has since increased. It is now devoted to establishing a comprehensive set of shared QC guidelines, tools for their capture, and microscopy metadata specifications for their storage and automated reporting.
Direct and open engagement of the global imaging community, and public and effective dissemination of QUAREP-LiMi advances, are essential to foster consensus-building and global acceptance of QUAREP-LiMi proposals. Consistent with this goal and its foundational principles, QUAREP-LiMi actively seeks participation from any interested party to join the group and work towards establishing a global QC consensus. To this aim, regular updates on the progress achieved by individual working groups will be disseminated using various social platforms and the QUAREP-LiMi website for all interested parties to provide input.

## Organizational Structure and Immediate Goals

Having established the general goals and founding principles of the group, members of QUAREP-LiMi agreed upon a set of essential topics to address and established the following organizational structureon July 9th, 2020:

1. The group aims to achieve specific deliverables and established that work will be conducted by individual working groups, each led by an elected Chair and Vice-Chair.
2. All interested parties are welcome to participate, either as observing members or as active participants within one or more working groups of their choice.
3. All participants, regardless of whether they are observing or active participants, will be allowed to provide feedback on the deliverables produced by individual working groups.
4. The number of working groups will be extended as needed to ensure coverage of all aspects of microscopy QC and satisfy all key beneficiaries' desiderata.

More specifically, the following working groups were established, and when appropriate, will produce a robust, easy-to-use protocol based upon standardized samples or tools:

***WG1 Illumination Power***

Comparison of fluorescence intensities between images requires measurements of the illumination power and stability of the excitation light source. WG1 aims at establishing a recommended protocol for measuring the stability of a light source during both short- and long-term image acquisition sessions using calibrated external power sensors. This initial aim will be extended later to measure the absolute flux of light through the illumination path and irradiance of the sample. The initial protocol will be designed around lasers on confocal microscopy platforms (raster scanning and spinning disks). It will be later modified towards other microscopy techniques (widefield, TIRF, light-sheet, super-resolution).

***WG2 Detection System Performance***

WG2 focuses on the detection system, comprising the detection path and its detector(s), and how it measures the signal from the sample. Members of WG2 aim to standardize the characterization of the detection system performance and create standard procedures for monitoring it over time, thereby revealing performance issues that could affect data reproducibility. Therefore, WG2 will define universal, externally measurable parameters applicable to any type of detector (e.g., photons, linearity, noise), together with measurement tools and protocols for measuring these parameters from common detector types. These universal parameters will be specified according to each distinct type of detector's internal parameters, which have already been defined by the community. They will enable the evaluation and comparison of different detection systems, thus pinpointing the most suitable technology for given applications.

***WG3 Uniformity of Illumination Field - Flatness***

Illumination field uniformity is critical for quantitative imaging when comparing fluorescence intensities across a field-of-view (FOV) or a large tile of images capturing an entire sample. If the illumination is not constant over a large area, the fluorescence intensities will not represent the inherent fluorescence but rather the location within the image. Thus, WG3 aims at defining a set of universal protocols to assess the uniformity of illumination (i.e., "flatness" of field) over the FOV of any photon-based imaging system and allow for correction of any non-uniformity. These protocols will identify the necessary tools, the procedures required to perform the measurements, and the analysis methods required for their interpretation. WG3 will also define criteria regarding the cut-off for acceptability and the need for correction. A database will be created with ideal images of uniform fields-of-view from different microscope modalities and settings to be used as a reference by the community and validate the protocol and criteria.

***WG4 System Chromatic Aberration and Co-Registration***

Chromatic aberration refers to possible artifacts caused by the wavelength dependency of an imaging system's optical properties, with the result that two colors arising from the same physical location within the sample appear separated in the image. Such artifacts result from the optical design of the system (e.g., well-corrected versus poorly corrected objective lenses), the manufacturing tolerances of the system components, and the alignment of the optical components. Co-registration accuracy more generally refers to the system's ability to co-localize dyes of different wavelengths emitting from the same object within a particular experimental set-up. This can be affected by both the experimental set-up and the system architecture. Working within the assumption that microscope users are ultimately interested in co-registration accuracy, WG4 aims to use sub-resolution and larger multi-colored bead preparations to measure co-registration accuracy. Alternative tools for performing these measurements will also be evaluated. WG4 will compare reproducibility across different laboratories to determine the best protocol.

***WG5 Lateral and Axial Resolution***

This WG focuses on the microscope lateral and axial resolution, which is essential for reporting size measurements of near-resolution limit objects or distances between them. Resolution is highly related to the objective quality but depends strongly upon other parameters ranging from the sample preparation to the signal detection.

The WG aims to define sample preparation, image acquisition, and data analysis protocols for testing resolution, first using sub-resolution fluorescent bead preparations and second employing alternative pattern-based methods. Monitoring the resolution (Point Spread Function in the case of beads) over time will identify possible aberrations in the system. Pooling the data from multiple laboratories within the WG will allow them to compare reproducibility for sample preparation, data acquisition, and data analysis tools, thereby determining a robust, easy-to-use protocol to propose to the community.

***WG6 Stage and Focus - Precision and Other***

The mission of WG6 is to ensure the performance and QC of stage platforms and sample holders and the optomechanical focus of the optical system as it relates to X, Y, Z movement, stability, reproducibility, and repeatability. The goals are defining the terms typically used to address QC, providing standardization of the measurements and testing protocols, and establishing performance benchmarking levels. Though initially applying these towards confocal light microscopy, the WG will endeavor to include details for standard incident light fluorescence microscopy and more advanced techniques such as super-resolution and light-sheet microscopy.

### WG7 Microscopy Data Provenance and QC Metadata

For proper interpretation, microscopy images must be accompanied by both human-readable (i.e., Materials and Methods sections) and machine-readable (i.e., metadata) descriptions of all steps leading to image formation (i.e., 'data provenance' metadata) as well as by QC metrics detailing the illumination, detection, chromatic, optical and mechanical performance of the microscope. Nevertheless, no universally accepted community guidelines exist defining what 'data provenance' and QC metadata should be reported for distinct types of imaging data. Therefore, the metadata automatically recorded by different commercial microscopes can vary widely, posing a substantial challenge for microscope users to create a *bona fide* record of their work. To meet these challenges, the 4D Nucleome (4DN)[41] Imaging Working Group and the BINA QC-DM WG[r] have developed a tiered set of Microscopy Metadata guidelines and a suite of extensions of the OME Data Model[12] that scale with experimental complexity and requirements, and are specifically tailored at enhancing comparability and reproducibility in light microscopy. WG7 aims to systematically evaluate the structure and semantics of the initial 4DN-BINA-OME extension proposal[10,11] and to launch a coordinated outreach strategy towards reaching a wide community consensus around the proposed metadata specifications.

### WG8 White Paper

The remit of WG8 is to relay both short and long-term goals of QUAREP-LiMi by the publication of a set of White Papers to communicate and seek cooperation from the community. The principal aim of these White Papers is to promote QUAREP-LiMi to 1) Prospective new members: to actively engage with the work of QUAREP-LiMi; 2) Imaging scientists and bioimage analysts: to raise awareness of QC issues; 3) Group Leaders/Principal Investigators: to engage a critical mass of academic researchers (top-down); 4) Research scientists (graduate students and postdoctoral researchers) with expertise in the specialized WG topics and imaging scientists: to influence the research group leaders (bottom-up); 5) Scientific publishers: to raise the quality of methods reporting and rigor and reproducibility in publications; 6) Leads (CEO/directors) of companies and commercial application specialists: to work alongside QUAREP-LiMi to facilitate ease of measurements and reporting, and 7) Prospective funders (funding agencies, private sponsors): to support the work of this initiative.

### WG9 Overall Planning and Funding

The principal aim of WG9 is to coordinate and promote the activities of QUAREP-LiMi. Within this WG, there is representation from all other WGs in addition to key global, regional, and national microscopy communities. WG9 will also liaise directly with corporate partners, scientific publishers, and funding bodies.

WG9 will focus on the following activities: 1) Ensure that the output of QUAREP-LiMi achieves maximum impact within the imaging community by raising awareness of the need for QC across all stakeholders in light microscopy (via white paper, website, publications); 2) Seek to obtain support from our corporate partners (microscope manufacturers/technology companies); 3) Obtain funding and support from national bodies, scientific publishers and learned societies to help cover the activities of QUAREP-LiMi (allow us to stage physical meetings, cover publication costs, help with organization and add impact); 4) Keep stakeholders informed and share information through a regularly updated website and tools database (towards internal and external communication and impact), and 5) Coordinate QUAREP-LiMi WGs and future QUAREP-LiMi meetings (virtual and physical).

### WG10 Image Quality

Good image quality (IQ) is essential for any subsequent image processing, analysis, and presentation steps. However, the notion of IQ is very broad and encompasses concepts that might differ between various microscope types. The aims of WG10 are 1) to define a set of basic IQ parameters (quantitative criteria, metadata, QC metrics) for light microscopy; 2) to weight the significance of the individual parameters for different experimental techniques and microscope types; and 3) to facilitate the assignment of a microscope- and experiment-specific QC rating to individual images. Ultimately, WG10 will work to summarize the upshot of these steps in the form of easy-to-use workflows. The integration of IQ ratings as part of the metadata associated with every imaging dataset is a long-term goal of this WG.

### WG11 Microscopy Publication Standards

WG11 will work together with scientific publishers to promote the adoption of best practices in the reporting of metadata (for both image acquisition and analysis) throughout scientific journals and books. Only by ensuring all relevant constituents (researchers and imaging scientists submitting publications and designing research; editors, scientific publishers, and reviewers monitoring and preparing publications; and funders, researchers, and educators evaluating and disseminating publications) are working in concert can we raise the bar to ensure reproducibility in imaging experiments. WG11 will focus on the following activities: 1) inform scientific publishers of the standards and metadata put forward by the other QUAREP Working Groups; 2) liaise with and encourage individual journals to modify their imaging guidelines to align with these recommendations; 3) work together with the scientific publishers to enforce high standards of imaging metadata reporting in all research works accepted for publication; 4) facilitate the involvement of technical reviewers with significant microscopy expertise during the review of papers that rely heavily on imaging techniques; 5) work together with publishers to promote and increase the appropriate acknowledgement and co-authorship of imaging scientists and core imaging facilities in publications; 6) encourage publishers to compel authors to make raw imaging data available if, and when, required for validation of published research and to make reasonable suggestions regarding duration of storage of raw imaging data relevant to published results; 7) propose minimum standards for figure quality, figure colour selection, scale bars, inserts, annotations and labelling, in order to render all microscopy figures easily interpretable by experts and non-experts alike.

---

[r] https://www.bioimagingna.org/qc-dm-wg

## Future Steps and Perspectives

The ultimate goal of QUAREP-LiMi is to benefit everybody in the light microscopy community. Our future strategy can be subdivided into medium-term goals to be achieved within the next few months and long-term goals to be realized within the next 1-2 years.

### *Medium-term goals*

*Growth and diversification of QUAREP-LiMi member body:* The vast majority of current QUAREP-LiMi members are imaging scientists representing academic labs, core imaging facilities, and standardization organizations. For the mission of QUAREP-LiMi to be successful, it is imperative to achieve greater engagement with industry (currently 17 companies with 25% of the total members), scientific publishers, funding agencies, and commercial microscope manufacturers. Moreover, a high priority is to achieve a better worldwide representation of the imaging community by including more members outside North America and Europe.

*Establishing a consensus of accepted guidelines within the WGs:* Each WG working towards a defined QC method and overall microscopy metadata specification will finalize a proposed solution and methodology for their topic. They will present this to the entire QUAREP-LiMi community (see “Organizational Structure and Immediate Goals” section) for evaluation, including some beta-testing. This will result in revised versions of individual WGs’ proposals submitted for final approval by the imaging community. The final approval of guidelines and methods will be a community decision and result in documented and openly accessible QC protocols. This kind of workflow is adopted and slightly modified from the proven workflow of the ISO.

### *Long-term goals*

*Dissemination of new guidelines to the scientific community and its stakeholders:* The QUAREP-LiMi guidelines will be published like those for RNAseq, proteomics, microarrays, etc.[3,42–44], and highlighted at national and international scientific meetings. Furthermore, the inclusion of the guidelines in teaching materials and training courses (both for microscope users and for imaging facility staff[45], as well as for commercial developers and corporate partners) will ensure their wide-spread adoption among microscope users and developers. The QUAREP-LiMi guidelines, initially developed for widefield and confocal microscopy, are intended to be adopted and extended to other imaging modalities, such as light-sheet and super-resolution microscopy.

*Implementation of the new guidelines within the community:* By engaging the entire imaging community throughout the development of new guidelines and specifications, we strive to implement a standard procedure for end-users and to promote the integration of the guidelines into commercial microscope hardware and software. As this process becomes easier and more streamlined, the long-term aim is to enable the automatic measurement of these metrics. Thus, early engagement with commercial manufacturers and other developers is critical to ensure simple approaches towards acquiring QC data.

*Working with stakeholders to promote the implementation of new guidelines:* A straightforward solution to encourage the uptake of minimal QC metrics would be for scientific publishers to adopt these standards as part of their standard requirements to accept material for publication. Ideally, access to raw data should also be provided. Such initiatives are gaining ground and being backed at the national level by funding bodies. Databases to store the QC data for each microscope should enable creating simple reports to accompany published experimental results, demonstrating the system’s real-time performance across the data collection period. In other fields (genomics, transcriptomics, proteomics, etc.), public data repositories have played a key role in implementing community-proposed standards and accelerating their adoption. Data repositories and journals worked together as de facto enforcers of a standard. Journals required data related to peer-reviewed manuscripts to be submitted to repositories. In contrast, repositories themselves enforce standards either at the time of submission or by converting submitted data to a standardized format for publication and download. With the establishment of several bioimage data publication systems[15,16,46,47], there is now an opportunity within the field of light microscopy to use a similar approach based on these successes.

*Modification of the existing ISO and establishing of new ISO standards based on guidelines developed by QUAREP-LiMi:* The final formalization of the QUAREP-LiMi guidelines will be achieved by their inclusion in new editions of the respective ISOs[18–20,48]. Whilst we expect that microscope QC will be a constantly evolving area, as new technologies become mainstream, the establishment of fixed versioning of the current guidelines for widefield and confocal systems will provide the community with a strong baseline for further developments of ISOs. It will cover the vast majority of current microscopy-based research.

## Conclusion

The international nature, size, and breadth of QUAREP-LiMi is critical for its mission, which will only succeed with sufficient buy-in from all stakeholders. The first step will be to reach a consensus between microscope and system component manufacturers, users, and microscope custodians regarding precisely what needs to be measured, how, and at what frequency, taking into account the experiment being performed and the downstream image analysis strategy. Next, a set of common, practical tools to accomplish these measurements must be developed and provided to the entire community. The microscope manufacturers can provide some of these as internal QC tools that align with the QUAREP-LiMi guidelines, thereby facilitating rapid, simple measurements by all microscope users and custodians. Such tools would help the companies ensure their instrumentation's consistent performance, facilitate the more rapid diagnosis of problems, and permit imaging scientists to perform necessary checks and alignments that must currently be performed by dedicated service engineers. Commercial microscope manufacturers can also support this paradigm shift by raising awareness of the importance of imaging QC with their direct customers; thus, their involvement in the QUAREP-LiMi initiative and working groups is crucial.

An equally critical factor in the agreed-upon guidelines' global adoption will involve education and raising awareness through publications, workshops, and meeting presentations explicitly targeted at research and imaging scientists. Since a significant part of many imaging scientists' responsibilities already lies in maintaining instrumentation at optimal performance, they are typically more familiar with the problems and challenges outlined above. Hence, they have tremendous potential to educate researchers on the importance of imaging QC, the tools available and recommended, and to disseminate the QUAREP-LiMi guidelines to their facility users and researchers who have microscopes in their laboratories. Funding bodies and scientific publishers could also encourage adopting these guidelines by requiring their implementation in all imaging-focused research. Scientific publishers can further educate their reviewers in imaging QC or bring in expert technical assessors to interrogate data quality and reliability. Publishers and reviewers should also encourage the sharing of imaging data in public repositories. Finally, repository hosts could help enforcement by automating data quality validation and ensuring that the data is made widely available to the broader scientific community.

QC is costly and requires significant time and effort, but its lack undermines trust in the quality of data, equipment, scientific rigor, reproducibility, and data exchange. By providing a clear community-driven way forward and working closely with all stakeholders, QUAREP-LiMi has the potential to drive a culture change. This will benefit the entire community by fundamentally transforming image data quality and reproducibility.

## Acknowledgments

We thank somersault18:24 BV (Leuven, Belgium) for help with Figure 1. E. C.-S. was supported by the project PPBI-POCI-01-0145-FEDER-022122, in the scope of Fundação para a Ciência e Tecnologia, Portugal (FCT) National Roadmap of Research Infrastructures.

## Author contributions

U.B. and G.N. coordinated the QUAREP-LiMi White Paper working group (WG8) and the effort to write this manuscript. All active members of the WG8 contributed significantly to the realization of the manuscript (from conceptual development and writing to the creation of the figures and glossary). R.N. coordinates the QUAREP-LiMi initiative.

## Competing interests

R.R.B., R.H., D.J.M., C.D.W., S.B., J.B., K.A.B., J.B., R.D., F.E., A.F., W.I.G., G.A.H., C.B.J., S.C.J., J.L., E.R., A.R., V.S., S.S., and D.S. have a financial interest as employees of companies producing, selling or distributing light microscopes or products related or used with light microscopy equipment. A.D.C. has a financial interest in PSFcheck and PyCalibrate.

# Appendix

## A1 Glossary

| | |
|---|---|
| Benchmarking | The comparison of device performance to an accepted standard for a given task. |
| Corporate Partners | Umbrella term including vendors and companies. We chose to use the word "partners" to encourage them to work with QUAREP-LiMi. |
| Data Management | Data Management comprises all disciplines related to managing data as a valuable resource. In particular, image data management can be defined as a process that includes all phases of the image data life cycle, from its ’pre-life’ to its ’after-life’, including experimental procedures, sample preparation, image acquisition, quality control, data-storage, -protections, -finding, -validation, -processing, -analysis, -interpretation, -sharing, -integration, presentation, re-use and archiving. As such, the ultimate goal of Data Management is to ensure the accessibility, reliability, reproducibility, and timeliness of image data for its users. In this context, it is important to note that the "FAIR Guiding Principles for scientific data management and stewardship" have been established to assess the quality of data management for scientific data. FAIR data are data that meet principles of findability, accessibility, interoperability, and reusability[39]. |
| Device Monitoring | The process of overseeing all activities and tasks that must be taken into account to adhere to a given level of standard performance. |
| Facility Microscope User | People using microscopes housed in core facilities; (Note: a student or postdoc could be BOTH a facility and non-facility user – the term used would depend on the given context). |
| Good Laboratory Practice | The Principles of Good Laboratory Practice are a managerial quality control system covering the organizational process and the conditions under which non-clinical health and environmental studies are planned, performed, monitored, recorded, reported, and retained (or archived). |
| Image Metadata | Metadata is data that provides information about other data. In particular, Image Metadata consists of any and all information that allows imaging results to be evaluated, interpreted, reproduced, found, cited, and re-used as established by measurable data quality criteria (i.e., FAIR principles[39]). As such, Image Metadata includes metadata that documents all phases of a typical bioimaging experiment including experimental treatment, sample preparation and labeling, image acquisition, instrument performance and image quality (i.e. Microscopy Metadata), and, last but not least, image processing and analysis (NOTE: As defined Image Metadata includes Microscopy Metadata). |
| Imaging Facility Staff | An umbrella term including directors, managers, staff scientists and technicians working in a recognized facility. |
| Imaging Scientists | An umbrella term for scientists whose job it is to make discoveries with imaging possible. They are typically trained across multiple disciplines and utilize a wide breadth of imaging technologies, (i.e., image data acquisition, analysis, and management) for the furtherance of science. An important role of imaging scientists is to interface with others using and developing this technology. For more information: http://www.imagingscientist.com/ |
| International Organization for Standardization (ISO) | The International Organization for Standardization (ISO) is an independent, non-governmental organization made up of members from the national standards bodies of 165 countries. ISO issues standards, such as *ISO 21073:2019 Microscopes - Confocal microscopes - Optical data of fluorescence confocal microscopes for biological imaging*[20]. |
| Microscopy Metadata | Metadata is data that provides information about other data. In particular, Microscopy Metadata is metadata that documents the process of Image Acquisition using a Microscope and the Quality of the resulting Image Data. As such, Microscope Metadata can be subdivided into two sub-categories: 'data provenance' metadata includes microscope hardware specifications and image acquisition settings; while quality control metadata includes calibration metrics that quantitatively assess the performance of the microscope at the time of acquisition and therefore allow the evaluation of image quality (NOTE: As defined, Microscopy Metadata is a sub-categoty of Image Metadata). |
| Non-Facility Microscope User | A researcher who uses their own microscope(s) in their own labs – or "Individual research groups operating microscopes" as a collective term. |
| Quality Assessment (QA) | Action performed to ensure the quality of a specific factor involved in image acquisition and analysis. |
| Quality Audit | Quality audit is the process of systematic examination of the quality system used to ensure image acquisition and analysis. It might be performed by internal or external experts, and ideally is comparing system quality over time to identify any issues. |
| Quality Control (QC) | Procedure performed to ensure the quality of all factors involved in image acquisition and analysis. |
| Quality Standards | Accepted level of device performance for a given set of tasks as defined and published by a recognized organization such as the ISO or DIN. |
| Standard Operating Procedures (SOPs) | Established methods to be followed routinely for the performance of specific operations. |
| Tools | Instruments, samples, or software used to determine the quality of all factors involved in image acquisition and analysis. |

*Last updated January 27, 2021.*